\documentclass[twocolumn]{article}

\usepackage{tikz}
\usepackage{amsmath}
\usepackage{amsmath,amsfonts}
\usepackage{algorithmic}
\usepackage{graphicx}
\usepackage{textcomp}
\usepackage{url}
\usepackage{hyperref}
\usepackage{multirow}
\usepackage{comment}
\usepackage{listings}
\usepackage{appendix}
\usepackage{stfloats}
\usepackage{caption} 
\usepackage{adjustbox} 
\definecolor{codegreen}{rgb}{0,0.6,0}
\definecolor{codegray}{rgb}{0.5,0.5,0.5}
\definecolor{codepurple}{rgb}{0.58,0,0.82}
\definecolor{backcolour}{rgb}{0.95,0.95,0.92}
\definecolor{lcol}{HTML}{AF3235}

\hypersetup{
    colorlinks=true,
    allcolors=lcol,
    linkcolor=black
}

\newcommand{\websiteurl}{https://sites.google.com/view/llm4cve}
\newcommand{\websitelink}[1]{\href{\websiteurl}{#1}}
\newcommand{\websitefn}{{\color{black}\footnote{\websiteurl}}}

\lstdefinestyle{mystyle}{
    language=c,
    backgroundcolor=\color{backcolour},   
    commentstyle=\color{codegreen},
    keywordstyle=\color{magenta},
    numberstyle=\tiny\color{codegray},
    stringstyle=\color{codepurple},
    basicstyle=\ttfamily\footnotesize,
    breakatwhitespace=false,         
    breaklines=true,                 
    captionpos=b,                    
    keepspaces=true,                 
    numbers=left,                    
    numbersep=5pt,                  
    showspaces=false,                
    showstringspaces=false,
    showtabs=false,                  
    tabsize=2
}

\begin{document}

\title{LLM4CVE: Enabling Iterative Automated Vulnerability Repair with Large Language Models}

\author{
  Mohamad Fakih$^{1*}$\thanks{$^{*}$These authors contributed equally.},
  Rahul Dharmaji$^{1*}$\footnotemark[1],
  Halima Bouzidi$^1$, \\
  Gustavo Quiros Araya$^2$,
  Oluwatosin Ogundare$^2$,
  Mohammad Abdullah Al Faruque$^1$ \\
  \\
  \normalsize
  $^1$Dept. of Electrical Engineering and Computer Science, 
  University of California, Irvine, CA, USA \\
  \normalsize
  $^2$Siemens Technology, Princeton, NJ, USA \\
  \normalsize
  \{mhfakih, rdharmaj, hbouzidi, alfaruqu\}@uci.edu \\
  \{gustavo.quiros, tosin.ogundare\}@siemens.com
}

\date{} 

\maketitle

\begin{abstract}
Software vulnerabilities continue to be ubiquitous, even in the era of AI-powered code assistants, advanced static analysis tools, and the adoption of extensive testing frameworks. It has become apparent that we must not simply prevent these bugs, but also eliminate them in a quick, efficient manner. Yet, human code intervention is slow, costly, and can often lead to further security vulnerabilities, especially in legacy codebases. The advent of highly advanced Large Language Models (LLM) has opened up the possibility for many software defects to be patched automatically. We propose \textbf{LLM4CVE} -- an LLM-based iterative pipeline that robustly fixes vulnerable functions in real-world code with high accuracy. We examine our pipeline with State-of-the-Art LLMs, such as \textbf{GPT-3.5}, \textbf{GPT-4o}, \textbf{Llama 3 8B}, and \textbf{Llama 3 70B}. We achieve a human-verified quality score of \textbf{8.51/10} and an increase in ground-truth code similarity of \textbf{20\%} with Llama 3 70B. To promote further research in the area of LLM-based vulnerability repair, we publish our testing apparatus, fine-tuned weights, and experimental data on our \websitelink{website}\websitefn.
\end{abstract}

\section{Introduction}

Human developers are prone to costly mistakes when designing software systems. Often, these are not simple errors, but a fundamental misunderstanding of cybersecurity concepts~\cite{votipkadev}. These vulnerable programs are targets for criminals to steal credentials, assets, and other valuable items from end-users~\cite{changproduction}. Moreover, the frequency of these types of attacks is steadily increasing~\cite{bugecon3}. As a result, the ability to quickly and efficiently rectify software bugs has become more critical than ever before.

\begin{figure}[t!]
    \centering
    \includegraphics[width=\linewidth]{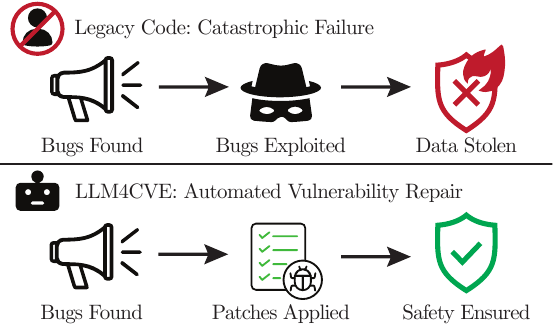}
    \caption{How LLM4CVE assists in preventing bug exploitation}
    \label{fig:timeline}
\end{figure}

Already, we have seen the significant impacts of these bugs -- billions of dollars in lost economic value~\cite{bugecon1, bugecon2, bugecon3}, countless man-hours spent on bug resolution~\cite{bughours1}, and the leakage of sensitive user data to malicious actors~\cite{bugimpact1, bugimpact2}. Cyberattacks -- like the one done to the company SolarWinds in 2020~\cite{solarwinds} -- are an example of this phenomenon. This attack infiltrated computers in both the U.S. government and private sector, leading to the leakage of millions of classified and confidential documents to foreign adversaries.

The proliferation of cyberattacks is not simply limited to one specific application domain.
The Internet-of-Things (IoT) revolution has led to the adoption of embedded systems in an ever-growing variety of devices~\cite{embsysforecast}. However, these systems are dangerously prone to critical security vulnerabilities~\cite{kimembsysbugs,chenembsysbugs,davidsonembsysbugs}. An attacker could feasibly extract personal information or confidential credentials by exploiting these bugs. Many such attacks have been deployed in the real world, leading to the proliferation of botnets~\cite{iotbotnet} and power grid disruptions~\cite{iotpowergrid}. As manufacturers race to incorporate new devices and features into their IoT products -- often without regard for the security and privacy of the end user -- these types of attacks are poised to proliferate in the future.

Even worse, autonomous vehicles are also vulnerable to many types of common security vulnerabilities \cite{avbugs}. These types of bugs are even more serious, as errors in control software can endanger the lives of pedestrians, passengers, or other drivers. While automatic testing suites have been created to verify the robustness of these autonomous systems in the physical world~\cite{avbugs2}, they are still prone to traditional software-driven attacks, much like any other safety-critical computer system. Therefore, these vehicles must be secured against vulnerabilities and bugs.

The shared link between the aforementioned vulnerable systems is their usage of common open-source software -- often written in C. These massive projects -- like the Linux kernel~\cite{linux}, OpenSSL~\cite{openssl}, and FFmpeg~\cite{ffmpeg} -- are frequently targeted by hackers. As a result, these large software projects contain a disproportionate amount of Common Vulnerabilities and Exposures (CVEs)~\cite{manycve}. A CVE informs the public about a known security vulnerability and serves as a centralized repository of information regarding the exploit~\cite{cve}. However, a CVE entry does not facilitate the automated repair of vulnerabilities on its own.

Several existing methods have been created to patch vulnerable software with minimal human input~\cite{pinvulnrev, zhangvulnrev, furepair, adamvulnrev, fuvision, harergan}. These techniques allow for the rectification of costly bugs with a comparatively lesser human cost. Yet, even these advanced systems are prone to mistakes. While these techniques often help rectify real-world bugs, automated vulnerability repair methods are known to generate invalid code~\cite{avrproblem1} or introduce additional bugs~\cite{avrproblem2}. More pressingly, these techniques often require analysis from a developer experienced with the program's codebase to fully implement~\cite{avrproblem3}. This presents challenges when security vulnerabilities are identified in old, unmaintained code -- or where the subject-matter experts for a particular product are no longer accessible. As a result, cybercriminals would be able to exploit these bugs, potentially leading to the leakage of sensitive user data. We demonstrate in Figure~\ref{fig:timeline} how our pipeline can mitigate the risks caused by abandoned or poorly maintained legacy code.

As an increasing amount of software governs critical real-world systems, the importance of program maintenance has grown drastically. The proportion of engineers devoted to maintaining legacy code systems has risen significantly~\cite{hoarelc}. Even then, the average time-to-fix of software vulnerabilities is only increasing~\cite{alexvulntime}. This presents a growing threat to end-users, especially when these bugs may take far longer to be patched in downstream code. 

The advent of highly capable Large Language Models (LLMs) has the potential to transform how software vulnerabilities are rectified, especially in older codebases. However, it is known that LLMs often produce flawed, uncompilable code~\cite{liucorrectness}. Even then, state-of-the-art LLMs such as GPT-4o~\cite{gpt4} and Llama 3~\cite{llama3} have spurred significant changes in software engineering practices. Moreover, specialized models tuned for code generation have appeared~\cite{codellama}, further increasing the potential for automated software augmentation and creation. Techniques such as Parameter-Efficient Fine-Tuning (PEFT)~\cite{peft} and Low-Rank Adaptations (LoRAs)~\cite{lora} extend the capabilities of these models, leading to an increase in performance while simultaneously streamlining the model training process~\cite{liu2022peft, xiongfinetune}. More recently, models incorporating a Mixture-of-Experts (MoE) have enabled significant gains in LLM performance~\cite{mistralmoe}. Researchers have also studied Prompt Engineering -- a method of refining LLM input to measurably improve the relevancy, accuracy, and quality of responses~\cite{prompteng1, prompteng2, prompteng3}. These advances in Large Language Models have created a unique opportunity for combination with existing vulnerability repair techniques to automatically rectify common software bugs.

We introduce LLM4CVE, an iterative pipeline that integrates Large Language Models with already-existing CVE data to fix common classes of software vulnerabilities with minimal human input. Given a snippet of code identified as faulty, our pipeline iterates until a viable candidate is obtained.
LLM4CVE begins generating a candidate fix, evaluating the viability of the fix, and applying required changes to increase the viability of the synthesized code. This iterative loop is further enhanced by the incorporation of LoRA fine-tuning into our pipeline. We also use prompt engineering to create a set of optimized prompts to enhance the quality of the generated code. Ultimately, LLM4CVE is capable of synthesizing a viable replacement code snippet, automating the vulnerability repair process for real-world examples of candidate CVEs. We summarize our key contributions as the following:

\begin{itemize}

    \item We present a novel, automated method for fixing security vulnerabilities in real-world programs that require minimal intervention from a skilled domain expert. Moreover, our approach is capable of preserving application security even in codebases with a small number of experienced maintainers.

    \item To the best of our knowledge, we create the first automated, iterative process for a Large Language Model to systematically correct vulnerabilities in code, improving on current automated vulnerability correction tools.

    \item We present a detailed study of the effectiveness of our iterative pipeline for fixing various classes of CVE/CWEs across multiple foundational models. Our methodology is tested on several mainstream LLMs -- including GPT-3.5, GPT-4o, Llama 3 8B, and Llama 3 70B.
    
\end{itemize}

The remainder of this paper is organized as follows: in Section \ref{sec:related}, we examine existing approaches to vulnerability repair -- including both manual and automated solutions. A discussion of the benefits of LLM4CVE over the current State-of-the-Art works follows. Next, we provide the reader with a brief background on both Large Language Models and automated vulnerability repair in Section~\ref{sec:bg}. We discuss the foundations of LLMs and the methods for fine-tuning them, as well as general methods of repairing software vulnerabilities without human intervention. A presentation of the LLM4CVE methodology follows in Section~\ref{sec:methods}, where each pipeline stage is thoroughly detailed. Our experimental setup and results are then displayed in Sections~\ref{sec:setup}~and~\ref{sec:results}. Finally, Sections~\ref{sec:discussion},~and~\ref{sec:conclusion} discuss the potential applications of our findings, known limitations and future improvements. The paper concludes with a public release of the LLM4CVE pipeline in Section~\ref{sec:code}.
\section{Background}
\label{sec:bg}

The LLM4CVE pipeline builds upon accepted wisdom in the field of Automated Vulnerability Repair, as well as incorporating state-of-the-art LLM augmentations -- including Parameter-Efficient Fine-Tuning (PEFT) and Low-Rank Adaptations (LoRAs) -- while also incorporating more traditional LLM techniques such as Prompt Engineering. In this section, we provide the reader with sufficient background knowledge for each domain, while also providing context for the design choices implemented in our pipeline.

\subsection{CVEs \& CWEs}
\label{sec:bgcve}

The reporting of bugs to centralized online repositories has become increasingly common. Entities such as the National Vulnerability Database~\cite{nvd} and MITRE's Common Vulnerabilities and Exposures~\cite{cve} provide up-to-date access to bug reports and mitigation measures. Often, when a security vulnerability is discovered in widely-used software, a description is listed on one or more of these databases~\cite{changproduction}. Since CVEs often target a specific software problem instead of describing a broad vulnerability class, the Common Weakness Enumeration (CWE) system was created to fulfill this purpose~\cite{householdercwe}. Ultimately, both the CVE and CWE identifiers are useful in our analysis for rectifying vulnerabilities.

\subsection{Vulnerability Analysis \& Repair}

The common goal of every vulnerability repair technique is to in some way rectify underlying software bugs that would otherwise lead to undefined or unsafe behavior. While formal models defining types of software deficiencies -- such as faults, errors and failures -- exist~\cite{bugconcepts}, we focus on the most simple definition -- the detection and repair of problems that result in unexpected output.

Automated vulnerability detection and repair often targets a specific class of bugs~\cite{monvulnrev, RAMPO}. This may be due to their heightened potential for exploitation, frequency in real-world code, or their ease of detection and subsequent patching, among other factors. These classes are commonly represented in CVEs and CWEs, as explained in Section~\ref{sec:bgcve}.

Common characterizations of vulnerability repair break the process into three sections -- (1) bug detection, (2) patching, and (3) patching~\cite{pinvulnrev}. The detection phase involves scanning source code using static analyzers~\cite{zhengstaticdetect}, machine learning~\cite{limldetect, chakramldetect}, or other methods~\cite{sotirovdetect}. Here, potentially faulty sections of code are identified for correction. Next, the deficient code is rectified through a variety of algorithmic~\cite{monvulnrev} and deep-learning~\cite{zhangvulnrev} methods. Before the patch is complete, a verification stage must confirm the reliability and robustness of the fix. This is often done through validation with a test suite~\cite{xiallmrepair3, avrproblem1}, static analysis~\cite{fangsanitize}, or by expert verification~\cite{gaorepair}. 

\begin{figure}[ht!]
    \centering
    \includegraphics[width=\linewidth]{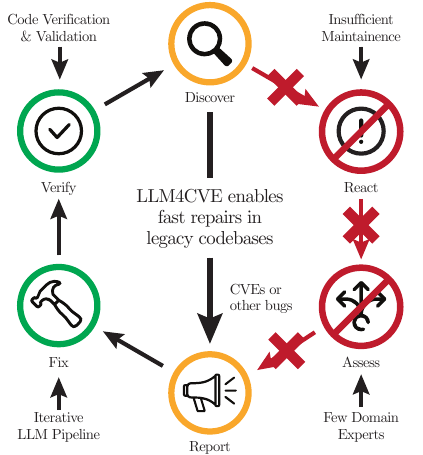}
    \caption{Rectification of software vulnerabilities often follows a predefined cycle, which LLM4CVE augments for faster turnaround times}
    \label{fig:vcycle}
\end{figure}

These methods are similar to industry-standard best practices regarding the software vulnerability life cycle~\cite{vcycle}.  An overview of these practices and how LLM4CVE is positioned in this cycle is shown in Figure~\ref{fig:vcycle}.

\subsection{Large Language Models}
Large Language Models (LLMs) leverage the attention mechanism~\cite{attention} to efficiently process complex inputs, capturing syntax and structure to understand dependencies between spatially distant but semantically linked tokens~\cite{shillmcontext, zaheercontext}. Methods like Retrieval-Augmented Generation further enhance these models by referencing external knowledge bases during inference to improve factual accuracy~\cite{lewisrag}.

The success of OpenAI's GPT series~\cite{gpt4} has fueled the rise of LLMs, with parameter counts growing exponentially—from GPT-3's 175 billion parameters~\cite{gpt3} to GPT-4's theorized 1.76 trillion~\cite{gpt4param}. This scaling increases performance and may lead to emergent abilities~\cite{llmemerge}, though the full impact is not yet fully understood~\cite{llmnoemerge}.

However, many leading LLMs, including those from OpenAI, are not open-source, posing challenges for fine-tuning and raising data privacy concerns. While OpenAI offers a fine-tuning API, it is restrictively licensed and costly~\cite{gptpricing}. Consequently, several open-source LLMs have emerged, such as Llama 3~\cite{llama3}, Code Llama~\cite{codellama}, and Mixtral (MoE)~\cite{mistralmoe}.

\subsubsection{LLM Augmentation}

To enhance LLM capabilities across diverse tasks, two key augmentation methods are used: (1) Parameter-Efficient Fine-Tuning/Low-Rank Adaptation (PEFT/LoRA), and (2) Mixture-of-Experts.

\begin{figure}[ht!] \centering \includegraphics[width=\linewidth]{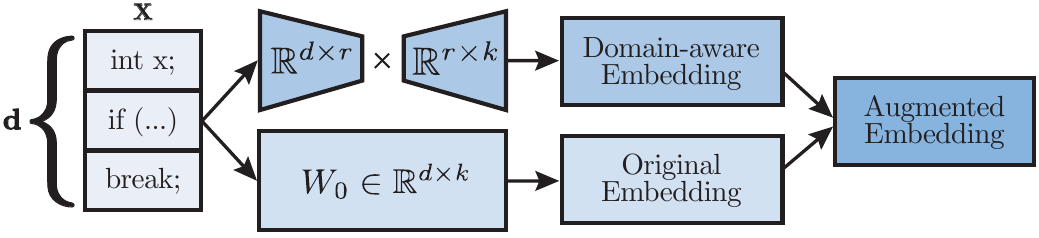} \caption{LoRAs enable the fine-tuning of LLMs with a comparatively low computational cost} \label{fig:lora} \end{figure}

The vast size of modern LLMs demands substantial computational resources for training. Parameter-Efficient Fine-Tuning (PEFT) addresses this by simplifying the fine-tuning process~\cite{zakenpeft, liupeft, fupeft}. A prominent PEFT method is Low-Rank Adaptation (LoRA), which freezes most model parameters and fine-tunes using injected rank decomposition matrices~\cite{lora}. Further improvements come from quantization~\cite{qlora} and applying LoRAs to Mixture-of-Experts models~\cite{loramoe}. A description of the LoRA training process is given in Figure~\ref{fig:lora}.

Another method is the Mixture-of-Experts (MoE) paradigm, which trains multiple specialized "expert" models and uses a router to select the optimal ones during inference~\cite{zhoumoe}. MoE models have demonstrated better scalability~\cite{dumoescale} and performance compared to compute-equivalent single-expert models~\cite{artetxemoeperf}.

\subsubsection{Prompt Engineering}

The quality of an LLM's output is directly influenced by the prompt quality. Refining prompts, known as \textit{Prompt Engineering}, improves performance on tasks like logic problems~\cite{prompteng3}, object annotation~\cite{shtedpe}, and general reasoning~\cite{whitepe}. Techniques such as Chain-of-Thought reasoning break down complex processes into individual steps for the LLM~\cite{chainofthought}. We employ a similar approach in the LLM4CVE pipeline, using automated compiler and metric-based feedback. LLMs can also perform zero-shot reasoning by being prompted to think step-by-step~\cite{kojimazeroshot}.

\begin{table*}[!ht]
\centering
\caption{An overview of selected CWEs for the LLM4CVE pipeline}
\begin{tabular}{|l|l|c|}
\hline
CWE & Title & Count \\
\hline
\hline
CWE-125 & Out-of-bounds Read & 452 \\
CWE-119 & Improper Restriction of Operations within the Bounds of a Memory Buffer & 363 \\
CWE-20  & Improper Input Validation & 289 \\
CWE-787 & Out-of-bounds Write & 179 \\
CWE-476 & NULL Pointer Dereference & 176 \\
CWE-190 & Integer Overflow or Wraparound & 156 \\
CWE-120 & Buffer Copy without Checking Size of Input ("Classic Buffer Overflow") & 121 \\
CWE-416 & Use After Free & 120 \\
\hline
\end{tabular}
\label{fig:cwetable}
\end{table*}

\section{Related Works} \label{sec:related}

Our work builds upon three primary areas: (1) automated vulnerability repair, (2) code generation with Large Language Models (LLMs), and (3) vulnerability detection and repair using LLMs. We integrate recent advancements in these fields to enhance the robustness and viability of our proposed pipeline, \textit{LLM4CVE}.

\subsection{Classical Automated Vulnerability Detection Repair}

Interest in automatically rectifying software vulnerabilities has been strong for decades. Traditional methods have leveraged program analysis tools, such as compilers or static analysis toolchains, to assist in fixing software bugs~\cite{zhangvulnrev,pinvulnrev,avrproblem1,avrproblem2,avrproblem3,monvulnrev,kliebercompiler}. However, static analysis has limitations in detecting certain types of bugs, such as those involving the Java Reflection API~\cite{landmanstatic}.

Advanced techniques have been developed to address these limitations. For example, methods using Generative Adversarial Networks (GANs) have been effective for vulnerability repair without the need for labeled training examples~\cite{harergan}. However, these methods are often evaluated only on synthetic code samples rather than real-world vulnerabilities.

Transfer learning has also been explored for automated software repair. The \textit{VRepair} framework demonstrated significant improvement over state-of-the-art methods, achieving almost a 50\% increase in repair rate using transformer architectures~\cite{chentransfer}. Despite its success, the model used is significantly smaller than modern LLMs like GPT-4~\cite{gpt4}.

Vision Transformers have been utilized to rectify code vulnerabilities by using special queries to locate vulnerable code snippets and generate accurate repair suggestions~\cite{fuvision}. This innovative model not only outperformed previous state-of-the-art models but was also reviewed positively by industry practitioners.

Code understanding models such as CodeT5~\cite{codet5} have enabled further improvements in the quality of generated fixes. As a precursor to modern LLMs like GPT-4 and Llama 3, the CodeT5 architecture allowed researchers to improve the total repair rate for software vulnerabilities. The proposed framework, \textit{VulRepair}, outperforms VRepair on several metrics due to extensive pre-training and the usage of Byte-Pair Encoding~\cite{furepair}.

Recent efforts aim to integrate multiple aspects of automatic programming to enhance effectiveness. Zhou \textit{et al.}~\cite{CREAM} proposed \textit{CREAM}, an automatic programming framework that leverages LLMs to integrate code search, code generation, and program repair. The framework retrieves similar code snippets through various code search strategies to guide the LLM's code generation process. It validates generated code using compilers and test cases, constructing repair prompts to query LLMs for correct patches. Preliminary experiments showed that \textit{CREAM} helped CodeLlama solve 267 programming problems with a 62.53\% improvement, demonstrating the potential of combining these three research areas.

Understanding why LLMs fail in code generation is crucial. Wen \textit{et al.}~\cite{llmfix} analyzed 12,837 code generation errors across 14 LLMs, identifying 19 distinct error causes. They proposed \textit{LlmFix}, a method that addresses three directly fixable error causes through a three-step process: correcting indentation, truncating redundant code, and importing missing modules. This approach significantly improved LLM performance on the HumanEval and MBPP datasets by 9.5\% and 5.4\%, respectively, highlighting the importance of automated error correction in enhancing code generation effectiveness.

\subsection{LLM-Driven Code Generation}

The popularization of LLMs has catalyzed significant interest in their use across various fields. Specifically, the success of automated code generation has been greatly accelerated by improvements in the logical reasoning abilities of these models~\cite{peillmreason}. State-of-the-art models like GPT-4~\cite{gpt4} have revolutionized code synthesis compared to their predecessors such as CodeBERT~\cite{codebert}.

Specialized LLMs for code synthesis have emerged, including CodeX~\cite{codex}, Code Llama~\cite{codellama}, WizardCoder~\cite{wizardcoder}, and CodeGen~\cite{codegen}. Many of these models are trained on publicly available software repositories, enhancing their code generation abilities. For example, CodeGen employs a multi-point synthesis scheme where the user is periodically prompted for feedback on the generated code~\cite{codegen}. However, this method requires active human intervention, as opposed to the fully automated feedback loop provided by \textit{LLM4CVE}.

Despite advancements, LLM-generated code often fails to pass test cases and requires human intervention. Prior efforts often neglected understanding why LLMs failed. Nguyen \textit{et al.}~\cite{safe} identified various error causes in LLM-generated code and showcased that directly addressing specific error types can significantly enhance the performance of multiple LLMs without modifying the models themselves.

Frameworks integrating testing and repair mechanisms have been proposed to improve code correctness. \textit{AutoSafeCoder}, introduced by Nunez \textit{et al.}~\cite{autosafecoder}, is a multi-agent framework that leverages LLM-driven agents for code generation, vulnerability analysis, and security enhancement through continuous collaboration. It consists of a Coding Agent, Static Analyzer Agent, and Fuzzing Agent, integrating dynamic and static testing iteratively during code generation to improve security. Experiments demonstrated a 13\% reduction in code vulnerabilities without compromising functionality.

\subsection{LLM-Guided Vulnerability Detection and Repair}

LLMs have been identified as key components for automatically rectifying software vulnerabilities. This is motivated by the ability of modern LLMs to generate consistently viable code snippets for many common programming languages~\cite{xullmeval}. Even for programming languages with relatively little toolchain support, augmented LLMs perform remarkably well~\cite{llm4plc}.

A significant amount of literature exists regarding the use of LLMs for automated program repair~\cite{yangllmrepair,silvallmrepair,zhoullmrepairreview,jinllmrepair,xiallmrepair,joshillmrepair,ahmedllmrepair}. Moreover, it has been demonstrated that LLM-based vulnerability repair pipelines can improve end-to-end repair rates~\cite{xiallmrepair2,xiallmrepair3,pearcezeroshot,wengllmrepair,jainllmrepair}.

One of the first works on this subject involves using zero-shot prompting to fix security vulnerabilities in a synthetic dataset~\cite{pearcezeroshot}. Since this work is from 2021, the LLMs used (Codex and Jurassic J-1) are relatively outdated. However, the performance of these techniques is already impressive, with a significant portion of simple, synthetic bugs fixed through the authors' pipeline.

Using the Codex and GPT-3 LLMs, researchers were able to repair 76.8\% of bugs in Java programs detected with static analysis tools~\cite{jinllmrepair}. Notably, these bugs were often security-related, falling into categories such as Null Pointer Dereferences and Thread Safety Violations. However, this tool is limited to the C\# and Java programming languages, whereas a majority of critical system software is written in languages such as C and C++~\cite{syslang,syslang2}.

There is an existing precedent for providing feedback in the LLM-driven code repair process. Existing implementations often require extensive test suites~\cite{xiallmrepair3}, which are not always available for real-world software. When given representative test cases, these frameworks can repair the majority of bugs automatically.

Nguyen \textit{et al.}~\cite{safe} proposed a novel framework that enhances the capability of LLMs to learn and utilize semantic and syntactic relationships from source code data for software vulnerability detection. Their \textit{SAFE} approach demonstrated superiority over state-of-the-art baselines on real-world datasets (ReVeal, D2A, Devign), achieving higher performances in F1-measure and Recall.

Researchers have also incorporated advanced LLM augmentation techniques such as Low-Rank Adaptations (LoRAs) to fine-tune their code repair models~\cite{silvallmrepair}. These methods are similar to our proposed \textit{LLM4CVE} pipeline and have shown significant improvements over non-augmented baselines. The datasets used in this work cover a wide variety of bugs present in everyday software rather than focusing specifically on security vulnerabilities. In comparison, we fine-tune our models on a dataset of real-world security vulnerabilities, as demonstrated in Section~\ref{sec:dataset}.
\section{Methodology}
\label{sec:methods}

\begin{figure*}[ht!]
    \centering
    \includegraphics[width=\textwidth]{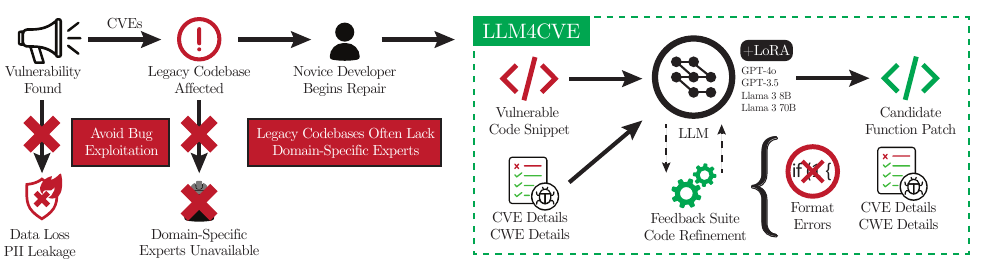}
    \caption{A visualization of how the LLM4CVE pipeline can automatically fix common software vulnerabilities}
    \label{fig:pipeline_overview}
\end{figure*}

LLM4CVE is an iterative pipeline that intends to automatically rectify common software vulnerabilities through the use of augmented Large Language Models. In this section, we aim to describe the structural and theoretical motivations behind the implementation of the pipeline.  A visualization of the LLM4CVE pipeline is given in Figure~\ref{fig:pipeline_overview}.

\subsection{CVE Selection}

We select eight of the most common CWEs for our analysis. A brief description of these CWEs along with their relative frequency in our dataset is provided in Table~\ref{fig:cwetable}. Details on the filtration of vulnerability examples for our pipeline are provided in Section~\ref{sec:datasetprep}.

\subsection{Prompting}

We employ both \textit{zero-shot} and \textit{one-shot} prompting for generating code snippets from an LLM. In \textit{one-shot} prompting, guidance is provided to the model over multiple rounds of code iteration, and the LLM is then able to incorporate feedback into the final code.
On the other hand, in \textit{zero-shot} prompting, no contextual guidance is provided to the model. Therefore, the LLM generates a code snippet without guidance or feedback. All LLMs in our study are tested with both types of prompting.

\subsection{Prompt Engineering}
\label{sec:prompteng}

The LLM4CVE pipeline employs two types of prompts, classified as "guided" and "unguided", respectively. The "guided" prompt provides the name of the CVE and CWE, a description of both, and specific instructions to further facilitate valid repair. The "unguided" prompt only instructs the model to repair the code -- no vulnerability details are provided. We also require the LLM to output code surrounded by delimiters to further optimize the extraction of code from the model output. Moreover, we instruct the model to create compilable code, use proper syntax and make the minimal amount of changes required to fix the target bug, as well as requiring it to synthesize all modifications itself without asking for user input.

Importantly, while CVE descriptions often provide a specific reference to a problem component in a function, CWE descriptions instead categorize the issue into a broad set of vulnerabilities. As a result, the "guided" prompt offers a more thorough description of the problem, which we posit allows for the LLM to generate superior candidate patches.

\subsection{Code Analysis}
\label{sec:codeanalysis}

To enable the automated evaluation of the LLM code output, an automated metric is required. For this purpose, we choose CodeBLEU, a widely accepted method for calculating the semantic similarity of two pieces of code~\cite{codebleu}. Like its predecessor BLEU~\cite{bleu}, which is often used for determining the quality of machine-translated text, CodeBLEU allows us to determine how similar the ground truth fixed code is to the LLM-generated fix. This tool provides scores in the range of 0-100 (we use an implementation that scales these values to 0-1), with a higher score implying the candidate code snippet is similar to the reference snippet. CodeBLEU incorporates the n-gram match from BLEU, in addition to in-depth analysis of code semantics via Dataflow Graphs and Abstract Syntax Trees.

\subsection{Iterated LLM Generation}
\label{sec:iterate}

One of the most important parts of the LLM4CVE pipeline is the automated feedback loop between analysis tools and the LLM. This allows for the LLM to generate improved code over multiple iterations, greatly increasing the quality of the final output. To implement this mechanism, we have used the difference in CodeBLEU scores between the broken input code and the LLM-generated output code.

The feedback provided conforms to two guiding principles -- (1) the output code should not be dramatically different than the input code, and (2) the output code should be valid C code. Our use of CodeBLEU scoring helps the model achieve both goals, as we can safeguard against excessive semantic changes to the code. Any faults found will automatically trigger another iteration of our pipeline. Then, if any new faults are found in the generated code (perhaps due to a failure of the model to synthesize a valid snippet), this process will be repeated. Ultimately, we collect output from both of these stages, along with the previously generated code and feed it into the LLM in the same context session for each iteration. We impose a limit of two iterations on the pipeline and select the second output as the candidate patch for evaluation.

\subsection{Evaluation Process}

Unlike many other approaches to automated vulnerability repair with Large Language Models, LLM4CVE targets real-world bugs instead of synthetic vulnerabilities. As a result, the evaluation of code correctness is a significantly more complicated problem. Many vulnerable code snippets are not self-contained, which means that external context is required to improve generation performance. We tailor of evaluation suite with this fact in mind, focusing on metrics highlighting improvements in code quality, and comparing the semantic similarity of our candidate patches to real-world ground truth solutions. In addition, we perform a selection of end-to-end compilation steps using our generated patches to confirm the viability of our pipeline. We believe this methodology offers a balance between theoretical evaluation and real-world applicability.

Importantly, we note that the nature of extracting function-level vulnerabilities from real-world codebases implies that these snippets are not fully self-contained. Therefore, it is infeasible to perform traditional static analysis, which necessitates our use of alternative techniques to generate feedback for the iterate steps of our pipeline.

There are multiple valid methods in which to fix a bug. For critical code, it is often more important for a vulnerability to be fixed immediately -- even if functionality is impacted. Therefore, a priority in our evaluation scheme is measuring the ability of our pipeline to fix the provided vulnerability. There are scenarios where the ground truth patch differs from our candidate patch, but we are concerned foremost with whether or not the bug is resolved. Then, both patches in this scenario would pass our evaluation method. 
\section{Experimental Setup}
\label{sec:setup}

In this section, we discuss the design of our experimental apparatus and provide details on the preprocessing, testing, and evaluation schemes of our work. It is important to note that we use multiple compute nodes equipped with one Nvidia A100, 48 CPU cores, and 256GB of system memory throughout our study.

\subsection{Datasets}
\label{sec:dataset}

Our primary dataset of interest is \textit{CVEFixes} -- a repository containing metadata, commit history, CVE/CWE classification, and most importantly: a before-and-after representation of vulnerable code~\cite{cvefixes}. This dataset contains over 10,000 vulnerable functions, a majority of which have labeled pairs of vulnerable (before) and non-vulnerable (after) code snippets. We use the provided SQL database to extract these labeled pairs, and we further filter them by language. We target the C programming language for the extraction of vulnerable code snippets.

\subsection{Dataset Preparation}
\label{sec:datasetprep}

As provided, the CVEFixes dataset does not lend itself to easy extraction of candidate before-and-after pairs corresponding to each CVE. Therefore, we implement a preprocessing pipeline to extract this information from the dataset. First, we obtain all function-level changes for the target programming languages ordered by CVE and function name, excluding CVEs with an associated CWE with minimal information such as "NVD-CWE-noinfo" and "NVD-CWE-other". We then filter out all CVE+name pairs that do not match our desired pattern of one vulnerable "before" code snippet, and one non-vulnerable "after" code snippet.

After obtaining a feasible set of before-and-after code snippets, we further trim these candidates by removing all pairs where at least one of the code snippets has a token count greater than 500. This ensures that we are not at risk of exceeding the context length of the Large Language Models used in our testing. After our dataset filtration stage, we are left with eight CWEs with at least 100 candidate pairs, representing 697 unique CVEs.

\subsection{Large Language Models}
\label{sec:llms}

For this study, we target the following Large Language Models: GPT-3.5, GPT-4o, Llama 3 8B, and Llama 3 70B. Notably, these models represent state-of-the-art performance in the categories of closed-source and open-source models. Importantly, the open-source nature of the Llama 3 family of LLMs enables the training of LoRAs to boost model performance, as explained in Section~\ref{sec:augment}. The range in parameter counts also offers an opportunity to explore the performance gradient between the selected models. It is also important to note that GPT-4o is a multimodal LLM, although our pipeline uses only text-based input.

Our selection of LLMs is also motivated by the maximum context length supported by each model. A longer context length enables the LLM to incorporate more information during the generation process, which is especially important during iterative generation, as each iteration builds on top of the existing context. Therefore, a sufficiently large context window is required for our pipeline to function, which is provided by all tested LLMs. This value was derived from the OpenAI documentation for GPT models~\cite{gptdocs} and the implementation specifications for the Llama 3~\cite{llama3} models. Note that the GPT-3.5-Turbo and GPT-4o models available from OpenAI represent the GPT-3.5 and GPT-4o models in our study. A description of the context length and parameter count for each model is provided in Table~\ref{fig:modelinfo}.

\begin{table}[ht!]
\centering
\caption{Context lengths and parameter counts of selected Large Language Models}
\begin{adjustbox}{width=\columnwidth}
\begin{tabular}{|c|c|c|}
\hline
Model & Context Length (Tokens) & Parameter Count\\
\hline
\hline
GPT-3.5 & 16,385 & 175B\\
\hline
GPT-4o & 128,000 & >1760B~\cite{gpt4param}\\
\hline
Llama 3 8B & 8,192 & 8B\\
\hline
Llama 3 70B & 8,192 & 70B\\
\hline
\end{tabular}
\end{adjustbox}
\label{fig:modelinfo}
\end{table}

\subsection{LLM Augmentation}
\label{sec:augment}

We train a Low-Rank Adaptation on the Llama 3 70B LLM using a portion of our created dataset. We employ an 90/10 train/test split to ensure sufficient data is available for our evaluation. We train
on labeled "broken"/"fixed" pairs, corresponding to the pre-fix and post-fix ground truth data. Then, we evaluate the LLM+LoRA on the test set by requesting for the set of broken code samples to be rectified. A complete description of our evaluation metrics is provided in Section~\ref{sec:metrics}.

\subsection{Pipeline Configurations}

We use three pipeline configurations -- (1) "unguided", (2) "guided", and (3) "guided+feedback". The third configuration also includes the trained LoRA for the Llama 3 70B model. An explanation of the prompting scheme and the feedback mechanism is shown in Section~\ref{sec:prompteng} and Section~\ref{sec:iterate}, respectively. Across our five tested models, this results in 15 potential model/configuration combinations. Note that we use a random sample consisting of 50\% of the full dataset for the "guided+feedback" configuration. We provide a full model/configuration diagram in Table~\ref{fig:configtable}.

\begin{table}[ht!]
\centering
\caption{Models \& pipeline configurations used in our study}

\begin{adjustbox}{width=\columnwidth}
\begin{tabular}{|c|c|c|c|}
\hline
Model & "unguided" & "guided" & "guided+feedback" \\
\hline
\hline
GPT-3.5 & $\checkmark$ & $\checkmark$ & $\checkmark$\\
\hline
GPT-4o & $\checkmark$ & $\checkmark$ & $\checkmark$\\
\hline
Llama 3 8B & $\checkmark$ & $\checkmark$ & $\checkmark$\\
\hline
Llama 3 70B & $\checkmark$ & $\checkmark$ & $\checkmark$\\
\hline
\end{tabular}
\end{adjustbox}
\label{fig:configtable}
\end{table}

\subsection{Automated Pipeline Feedback}
\label{sec:iterateimpl}

The LLM4CVE pipeline employs a "guided with feedback" approach that combines prompt engineering and iterative generation to produce higher-quality vulnerability patches, limited to two iterations for performance and computational efficiency.

To avoid exceeding the LLM's context window and prevent catastrophic forgetting, each prompt and vulnerable code snippet is kept around 500 tokens. Feedback prompts are similarly sized because they include the in-progress code at each iteration step.

After obtaining output from the LLM, and if we have not reached our iteration limit as defined in Section~\ref{sec:iterate}, we extract the code and submit it for testing. We analyze the change in CodeBLEU score between the current and previous code versions, as explained in Section~\ref{sec:codeanalysis}, to identify issues. If the CodeBLEU score diverges significantly, it likely indicates an error, and we inform the LLM that the candidate patch may be incorrect in the next prompt. Importantly, we never compare the CodeBLEU score of the candidate patch to the ground truth, as our pipeline would not have access to the ground truth fix in real-world use.

\begin{figure}[ht!]
    \centering
    \includegraphics[width=\linewidth]{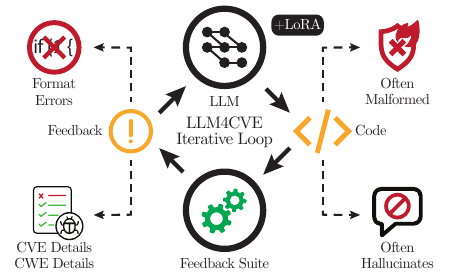}
    \caption{LLM4CVE uses iterative generation to improve the overall quality of patch synthesis}
    \label{fig:iterate}
\end{figure}

A complete description of our pipeline's iterative generation step is provided in Figure~\ref{fig:iterate}.

\subsection{Candidate Patch Extraction}
\label{sec:patchextract}

After obtaining an output from the LLM, we must extract only the code from it, discarding any delimiters or extraneous commentary. From our analysis, approximately 95\% of responses are well-formed, following the code block formatting specified in the prompt. For these responses, we simply extract the candidate patch from the code block. The remaining responses require nontrivial logic to extract patches from, which we implement as well. In the case where we detect no code has been generated (<1\% of samples), we output no candidate patch instead.

\subsection{Metrics}
\label{sec:metrics}

We use four principal metrics for evaluation -- CodeBLEU scores, human quality evaluation, end-to-end compilation, and required engineering effort. We employ a pass @ k scheme, with $k=1$, as described in~\cite{codex}. Detailed descriptions of our chosen metrics are provided below.

\subsubsection{CodeBLEU Scores}

We evaluate the final output code from all configurations and model types against the ground truth non-vulnerable code snippet. As a higher CodeBLEU score implies greater semantic similarity between two pieces of code, a large (i.e. near 1.00) score between the ground truth and the output code implies greater potential for valid bug correction. Importantly, a candidate patch with a CodeBLEU score less than 1.0 can still be a viable fix, as there are often multiple solutions to the given vulnerability.

\subsubsection{Human Quality Scores}

This metric provides a measure of the functionality and reliability of the proposed patch. A high score in this category implies that the generated fix is likely to be high-quality, usable code. Moreover, the demonstration of greater confidence in the proposed patch by the human reviewers also validates the practical applicability of our pipeline.

\subsubsection{End-to-End Compilation}

Next, we test our pipeline's generated patches in real-world codebases. We  directly apply the result of our pipeline to the codebase affected by the vulnerability. Then, we compile the entire project and determine the validity of the fix. This metric measures the ability for the LLM4CVE pipeline to fix real-world security vulnerabilities. It also ensures that the pipeline output is compliant, compiliable code.

\subsubsection{Engineering Effort}

Finally, we analyze the engineering effort required between traditional approaches and our pipeline. We compare setup times, the level of experience required, and the technical complexity of the proposed technique.
\section{Results}
\label{sec:results}

In this section, we present the results of our evaluation of the LLM4CVE pipeline. We measure four key metrics -- CodeBLEU Scores, Human Quality Scores, End-to-End Compilation success rate, and required Engineering Effort. These metrics enable a multifaceted assessment of the practicality, functionality, and effectiveness of the LLM4CVE pipeline. We provide visual comparisons between the various configuration of our pipeline over the five Large Language Models described in Section~\ref{sec:llms}. Moreover, we compare pipeline results between all three configurations across the eight CVEs used.

\subsection{CodeBLEU Scores}

For this metric, we compare the CodeBLEU scores between the pipeline output and the ground truth fix. Our CodeBLEU software tool generates semantic similarity ratings in the range of 0.0-1.0. A score of 1.0 implies an exact match when considering n-grams, syntax, and dataflow between the two samples. An important consideration to keep in mind is that there may be multiple "valid" fixes for a CVE, so an inexact match is not necessarily indicative of invalid code. Therefore, we treat this metric as a probabilistic estimate of the likelihood of generating a viable candidate fix. We extend this evaluation with real-world evaluation of selected candidate patches in Section~\ref{sec:e2e}.

\begin{figure}[ht!]
    \centering
    \includegraphics[width=\linewidth]{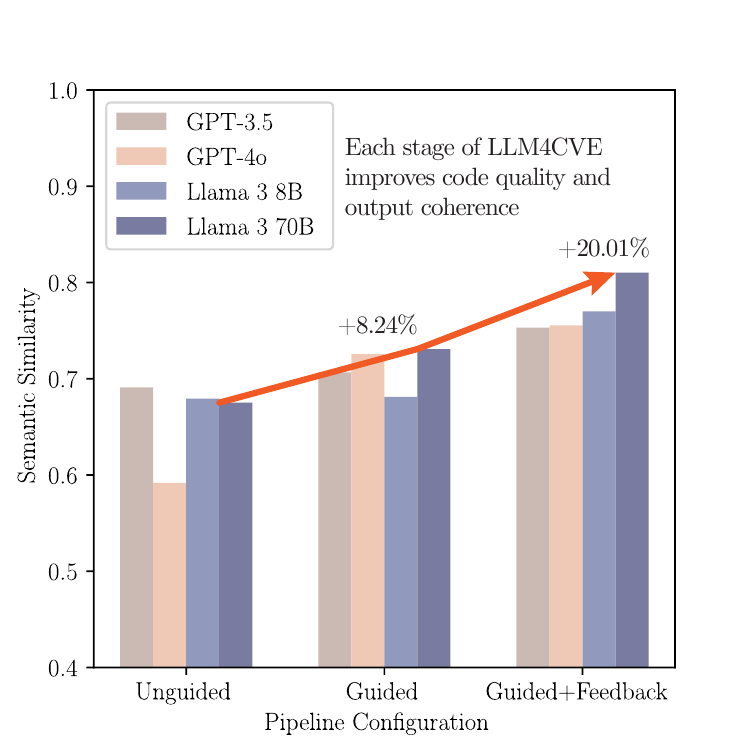}
    \caption{Semantic similarity scores across all pipeline configurations and model types}
    \label{fig:codebleu}
\end{figure}

We include results for all three pipeline configurations -- "unguided" (zero-shot), "guided" (one-shot), and "guided+feedback" (one-shot with feedback) -- and the semantic similarity scores are presented in Figure~\ref{fig:codebleu}. Importantly, the full configuration of the LLM4CVE pipeline -- "guided+feedback" -- demonstrates a remarkable performance improvement across all models, with the Llama 3 70B LLM peaking at a 20\% increase in semantic similarity scores.

\subsection{Human Quality Scores}

We perform a human study\footnote{We received prior approval to conduct this study from an institutional IRB through an exemption due to the strictly academic nature of our questionnaire.} over a group of participants with at least several years of experience in programming. Our metrics are centered around functionality evaluations. Participants understood the purpose of the study -- including the presence of one ground-truth patch and two LLM-generated patches per example -- but were not told which LLM created each candidate patch. In total, we include selected outputs from the Llama 3 70B and GPT-4o models. We include the output of the "guided+feedback" configuration for each model, along with the ground truth fixed code snippet, for three total candidate patches per function.

For each function, we showed participants the CWE name, CWE description, CVE name, and CVE description for the vulnerability in question. Then, we asked them the following questions:

\begin{itemize}
\item Vulnerability Elimination: On a scale of 1-10, how accurately does the following code eliminate the security vulnerabilities mentioned above?
\item Code Style: On scale of 1-10, how well does the following code adhere to best practices for code style?
\end{itemize}

Note that we compare and normalize the results of the first question with the known correctness of the ground-truth score in mind. We also observe that a majority of correctness ratings are centered around 1 (completely incorrect), or 10 (completely correct). Performing a qualitative evaluation of the LLM4CVE pipeline highlights its strengths in a real-world setting. Moreover, this approach allows for a multifaceted assessment of the usefulness and practicality of our LLM-based vulnerability repair model by industry practitioners. We present the results of the human study in Table~\ref{fig:humanstudy}.

\begin{table}[ht!]
\centering
\caption{Human evaluation of the LLM4CVE pipeline}

\begin{adjustbox}{width=\columnwidth}
\begin{tabular}{|c|c|c|c|}
\hline
Patch Source & Configuration & Style & Correctness \\
\hline
\hline
GPT-4o & "guided+feedback" & \textbf{9.01} & 7.44\\
\hline
Llama 3 70B & "guided+feedback" & 7.81 & \textbf{8.51}\\
\hline
Ground Truth & $\times$ & 4.24 & 10.00\\
\hline
\end{tabular}
\end{adjustbox}
\label{fig:humanstudy}
\end{table}
\subsection{End-to-End Compilation}
\label{sec:e2e}

For this test, we focus on evaluating the pipeline's effectiveness in fixing real-world bugs. We begin by initializing our pipeline with a vulnerable function and associated prompt. As an example, the \verb|cJSON_DeleteItemFromArray| function from the \verb|iperf3| tool is chosen. This tool is a TCP, UDP, and SCTP network bandwidth measurement tool, and uses the library \verb|cJSON|, in which this vulnerability lies. The software is affected by CVE-2016-4303 (CWE-120).

Once a candidate patch is obtained, we move to evaluation of its viability. We create a testing harness that exploits the vulnerability at hands and compile the program with and without the candidate patch. Our harness poses as a malicious actor trying to crash the software by providing a malformed \verb|cJSON| object with invalid lengths. Then, we run both programs and note which program crashes. Through this evaluation, it was confirmed that the candidate patch prevented the malicious actor from exploiting this vulnerability by rejecting the malformed object, demonstrating the effectiveness of the LLM4CVE pipeline. We provide further details of our testing harness in our public release of the pipeline, which can be found in Section~\ref{sec:code}.

\subsection{Engineering Effort}

Software solutions must be evaluated for efficiency and real-world practicality. We compare the ease of use and complexity of traditional human repair, GPT-based solutions, and open-source LLMs. Human repair costs are estimated from published data, while other statistics come directly from our pipeline.

GPT models are relatively efficient to set up due to OpenAI's simple API for inference, requiring no user compute resources. Open-source LLMs offer data security by allowing on-site training and inference but require significantly more setup time and manual configuration; their speed depends on access to GPU resources.

Trained engineers are the slowest method, as skilled human labor is needed to fix security vulnerabilities. In legacy codebases, this issue is worsened due to inaccessible or missing expertise. Studies estimate that human engineers take between several days~\cite{mobugtime} and one month~\cite{cvefixes} to repair a security vulnerability. We combine setup and execution time into one statistic for this category.

Time values for LLMs are based on our team's average time to set up a basic GPT/Llama pipeline and produce a single candidate patch. The LoRA training process requires access to relevant datasets like \textit{CVEFixes}, and we include an estimate of training time in our evaluation. A comparison of the average time costs among these approaches is given in Table~\ref{fig:timetable}.

\begin{table}[ht!]
\centering
\caption{Comparison of engineer-hours required for selected vulnerability patching techniques}

\begin{adjustbox}{width=\columnwidth}
\begin{tabular}{|c|c|c|}
\hline
Technique & Setup Time & Execution Time \\
\hline
\hline
Human Intervention & $\times$ & 28 days~\cite{cvefixes}\\
\hline
Open-Source LLMs+LoRAs & 24 hours & 10 minutes\\
\hline
GPT LLMs & 1 hour & \textbf{5 minutes}\\
\hline
\end{tabular}
\end{adjustbox}
\label{fig:timetable}
\end{table}

\section{Discussion}
\label{sec:discussion}

Our proposed pipeline lowers the barrier to entry for repairing critical security vulnerabilities, especially in legacy codebases. Significantly less engineering effort is required in these types of projects with deteriorating knowledge bases and few active maintainers. Moreover, the speed and efficiency of our pipeline enable these fixes to be made with haste, lessening the time between the discovery of a bug and its associated patch.

\subsection{Impact}

The LLM4CVE pipeline enables the quick repair of vulnerabilities in critical system software. Moreover, our tool is of increased relevance to legacy codebases, where devoted maintainers are often in short supply, and turnaround times are often long. By lessening the dependence on these domain-specific experts, our pipeline enables critical security vulnerabilities to be patched faster, increasing the overall safety of these programs. 

These benefits are further enhanced by the speed of our pipeline, as demonstrated in Table~\ref{fig:timetable}. Applying patches in time-sensitive environments -- such as when a new vulnerability is discovered -- is desirable to maintain system integrity and robustness.

\subsection{Candidate Patch Assessment}

It is of interest to compare a patch generated by the LLM4CVE pipeline to the original vulnerable function. We provide a candidate patch for the \verb|snmp_ber_decode_type| function affected by CWE-125 and CVE-2020-12141. Below, Listing~\ref{lst:vulnpatch} is the original vulnerable code, while Listing~\ref{lst:gptpatch} is generated by GPT-4o using our fully-featured pipeline.

\begin{lstlisting}[label=lst:vulnpatch, caption=The original vulnerable function]
unsigned char * snmp_ber_decode_type(unsigned char *buff, uint32_t *buff_len, uint8_t *type)
{
  if(*buff_len == 0) {
    return NULL;
  }

  *type = *buff++;
  (*buff_len)--;

  return buff;
}
\end{lstlisting}

\begin{lstlisting}[label=lst:gptpatch, caption=A candidate patch generated by GPT-4o]
unsigned char* snmp_ber_decode_type(unsigned char *buff, uint32_t *buff_len, uint8_t *type) {
  if(buff == NULL || buff_len == NULL || type == NULL || *buff_len == 0) {
    return NULL;
  }

  *type = *buff++;
  (*buff_len)--;

  return buff;
}
\end{lstlisting}

We see that the LLM output checks the validity of the input variables \verb|buff|, \verb|buff_len|, and \verb|type|, while the vulnerable code only attempts to validate \verb|buff_len| (and that too with a blind dereference). Then, it is evident that our pipeline was able to patch the vulnerability in a viable, non-destructive manner.

\subsection{GPT vs. Llama}

Throughout our experiments, the Llama 3 70B model consistently matched or outperformed other LLMs, especially using the full LLM4CVE pipeline. Even the relatively smaller Llama 3 8B model can compete with the GPT models once fine-tuning is performed. This demonstrates the effectiveness of Parameter-Efficient Fine-Tuning techniques like LoRAs, as without these adapters the GPT and Llama models perform roughly equivalently in the "guided" (one-shot) pipeline configuration. The largest gains in generation ability were derived from the iterative configuration of our pipeline, and this can be attributed to our fine-tuning of open-source models. Some LLMs such as GPT-4o are not fine-tunable, and so we are unable to apply all techniques mentioned in Section~\ref{sec:methods} to these models.

\subsection{Ethical Considerations}

Our tool serves to rectify vulnerabilities in codebases where the maintainers would otherwise be unable to do so themselves. As a result, we expect that: (1) public usage of our tool will serve the security community by keeping the end-user better protected from cybercriminals, and (2) there are no significant ethical risks posed by our pipeline. In addition, all vulnerabilities studied were already publicly disclosed by nature of being a known CVE. Therefore, no further vulnerability disclosure is necessary.

\section{Conclusion}
\label{sec:conclusion}

The LLM4CVE pipeline serves to fix security vulnerabilities in critical system software with minimal human input. By combining traditional bug repair methods with state-of-the-art Large Language Model techniques, we improve the robustness and viability of automated program repair. Our iterative pipeline allows for gradual refinement of the generated code, which increases the likelihood of obtaining a viable candidate patch. We further extend LLM4CVE with automated code analysis tools, LLM fine-tuning, and Prompt Engineering. A thorough evaluation of real-world vulnerabilities through both automated and human-centered means has shown the efficacy of our approach, and we believe our contributions to the field will pave the way towards achieving automated program repair without any intervention from trained experts.
\section{Code Availability}
\label{sec:code}

We publish our testing apparatus, fine-tuned weights, and experimental data on our \websitelink{website}\websitefn.

\bibliographystyle{unsrt}
\bibliography{llm4cve.bib}

\begin{thebibliography}{100}

\bibitem{votipkadev}
Daniel Votipka, Kelsey~R. Fulton, James Parker, Matthew Hou, Michelle~L. Mazurek, and Michael Hicks.
\newblock Understanding security mistakes developers make: Qualitative analysis from build it, break it, fix it.
\newblock In {\em 29th USENIX Security Symposium (USENIX Security 20)}, pages 109--126. USENIX Association, August 2020.

\bibitem{changproduction}
Yung-Yu Chang, Pavol Zavarsky, Ron Ruhl, and Dale Lindskog.
\newblock Trend analysis of the cve for software vulnerability management.
\newblock In {\em 2011 IEEE Third International Conference on Privacy, Security, Risk and Trust and 2011 IEEE Third International Conference on Social Computing}, pages 1290--1293, 2011.

\bibitem{bugecon3}
{Internet Crime Complaint Center}.
\newblock 2023 internet crime report.

\bibitem{bugecon1}
Amit Elazari~Bar On.
\newblock The law and economics of bug bounties, August 2018.

\bibitem{bugecon2}
Nicolas Huaman, Bennet von Skarczinski, Christian Stransky, Dominik Wermke, Yasemin Acar, Arne Drei{\ss}igacker, and Sascha Fahl.
\newblock A {Large-Scale} interview study on information security in and attacks against small and medium-sized enterprises.
\newblock In {\em 30th USENIX Security Symposium (USENIX Security 21)}, pages 1235--1252. USENIX Association, August 2021.

\bibitem{bughours1}
{Osterman Research, Inc}.
\newblock Second annual state of ransomware report: Survey results for australia.

\bibitem{bugimpact1}
Andreas Kogler, Jonas Juffinger, Lukas Giner, Lukas Gerlach, Martin Schwarzl, Michael Schwarz, Daniel Gruss, and Stefan Mangard.
\newblock {Collide+Power}: Leaking inaccessible data with software-based power side channels.
\newblock In {\em 32nd USENIX Security Symposium (USENIX Security 23)}, pages 7285--7302, Anaheim, CA, August 2023. USENIX Association.

\bibitem{bugimpact2}
Joel Reardon, {\'A}lvaro Feal, Primal Wijesekera, Amit Elazari~Bar On, Narseo Vallina-Rodriguez, and Serge Egelman.
\newblock 50 ways to leak your data: An exploration of apps{\textquoteright} circumvention of the android permissions system.
\newblock In {\em 28th USENIX Security Symposium (USENIX Security 19)}, pages 603--620, Santa Clara, CA, August 2019. USENIX Association.

\bibitem{solarwinds}
Sean Peisert, Bruce Schneier, Hamed Okhravi, Fabio Massacci, Terry Benzel, Carl Landwehr, Mohammad Mannan, Jelena Mirkovic, Atul Prakash, and James~Bret Michael.
\newblock Perspectives on the solarwinds incident.
\newblock {\em IEEE Security \& Privacy}, 19(2):7--13, 2021.

\bibitem{embsysforecast}
Shadi Al-Sarawi, Mohammed Anbar, Rosni Abdullah, and Ahmad~B. Al~Hawari.
\newblock Internet of things market analysis forecasts, 2020–2030.
\newblock In {\em 2020 Fourth World Conference on Smart Trends in Systems, Security and Sustainability (WorldS4)}, pages 449--453, 2020.

\bibitem{kimembsysbugs}
Taegyu Kim, Vireshwar Kumar, Junghwan Rhee, Jizhou Chen, Kyungtae Kim, Chung~Hwan Kim, Dongyan Xu, and Dave~(Jing) Tian.
\newblock {PASAN}: Detecting peripheral access concurrency bugs within {Bare-Metal} embedded applications.
\newblock In {\em 30th USENIX Security Symposium (USENIX Security 21)}, pages 249--266. USENIX Association, August 2021.

\bibitem{chenembsysbugs}
Libo Chen, Yanhao Wang, Quanpu Cai, Yunfan Zhan, Hong Hu, Jiaqi Linghu, Qinsheng Hou, Chao Zhang, Haixin Duan, and Zhi Xue.
\newblock Sharing more and checking less: Leveraging common input keywords to detect bugs in embedded systems.
\newblock In {\em 30th USENIX Security Symposium (USENIX Security 21)}, pages 303--319. USENIX Association, August 2021.

\bibitem{davidsonembsysbugs}
Drew Davidson, Benjamin Moench, Thomas Ristenpart, and Somesh Jha.
\newblock {FIE} on firmware: Finding vulnerabilities in embedded systems using symbolic execution.
\newblock In {\em 22nd USENIX Security Symposium (USENIX Security 13)}, pages 463--478, Washington, D.C., August 2013. USENIX Association.

\bibitem{iotbotnet}
Manos Antonakakis, Tim April, Michael Bailey, Matt Bernhard, Elie Bursztein, Jaime Cochran, Zakir Durumeric, J.~Alex Halderman, Luca Invernizzi, Michalis Kallitsis, Deepak Kumar, Chaz Lever, Zane Ma, Joshua Mason, Damian Menscher, Chad Seaman, Nick Sullivan, Kurt Thomas, and Yi~Zhou.
\newblock Understanding the mirai botnet.
\newblock In {\em 26th USENIX Security Symposium (USENIX Security 17)}, pages 1093--1110, Vancouver, BC, August 2017. USENIX Association.

\bibitem{iotpowergrid}
Saleh Soltan, Prateek Mittal, and H.~Vincent Poor.
\newblock {BlackIoT}: {IoT} botnet of high wattage devices can disrupt the power grid.
\newblock In {\em 27th USENIX Security Symposium (USENIX Security 18)}, pages 15--32, Baltimore, MD, August 2018. USENIX Association.

\bibitem{avbugs}
Garcia Joshua, Feng Yang, Shen Junjie, Almanee Sumaya, Xia Yuan, Chen, and Qi~Alfred.
\newblock A comprehensive study of autonomous vehicle bugs.
\newblock In {\em Proceedings of the ACM/IEEE 42nd international conference on software engineering}, pages 385--396, 2020.

\bibitem{avbugs2}
Seulbae Kim, Major Liu, Junghwan~"John" Rhee, Yuseok Jeon, Yonghwi Kwon, and Chung~Hwan Kim.
\newblock Drivefuzz: Discovering autonomous driving bugs through driving quality-guided fuzzing.
\newblock In {\em Proceedings of the 2022 ACM SIGSAC Conference on Computer and Communications Security}, CCS '22, page 1753–1767, New York, NY, USA, 2022. Association for Computing Machinery.

\bibitem{linux}
{Linux Foundation}.
\newblock Linux kernel.

\bibitem{openssl}
{The OpenSSL Project}.
\newblock Openssl.

\bibitem{ffmpeg}
{FFmpeg Team}.
\newblock Ffmpeg.

\bibitem{manycve}
Norman Santiago and Janelli Mendez.
\newblock Analysis of common vulnerabilities and exposures to produce security trends.
\newblock In {\em Proceedings of the 2022 International Conference on Cyber Security}, CSW '22, page 16–19, New York, NY, USA, 2023. Association for Computing Machinery.

\bibitem{cve}
MITRE Corporation.
\newblock {\em CVE - MITRE}.

\bibitem{pinvulnrev}
Eduard Pinconschi, Rui Abreu, and Pedro Adão.
\newblock A comparative study of automatic program repair techniques for security vulnerabilities.
\newblock In {\em 2021 IEEE 32nd International Symposium on Software Reliability Engineering (ISSRE)}, pages 196--207, 2021.

\bibitem{zhangvulnrev}
Quanjun Zhang, Chunrong Fang, Bowen Yu, Weisong Sun, Tongke Zhang, and Zhenyu Chen.
\newblock Pre-trained model-based automated software vulnerability repair: How far are we?
\newblock {\em IEEE Transactions on Dependable and Secure Computing}, pages 1--18, 2023.

\bibitem{furepair}
Michael Fu, Chakkrit Tantithamthavorn, Trung Le, Van Nguyen, and Dinh Phung.
\newblock Vulrepair: a t5-based automated software vulnerability repair.
\newblock In {\em Proceedings of the 30th ACM Joint European Software Engineering Conference and Symposium on the Foundations of Software Engineering}, ESEC/FSE 2022, page 935–947, New York, NY, USA, 2022. Association for Computing Machinery.

\bibitem{adamvulnrev}
Adam Doup\'{e}.
\newblock History and future of automated vulnerability analysis.
\newblock In {\em Proceedings of the 24th ACM Symposium on Access Control Models and Technologies}, SACMAT '19, page 147, New York, NY, USA, 2019. Association for Computing Machinery.

\bibitem{fuvision}
Michael Fu, Van Nguyen, Chakkrit Tantithamthavorn, Dinh Phung, and Trung Le.
\newblock Vision transformer inspired automated vulnerability repair.
\newblock {\em ACM Trans. Softw. Eng. Methodol.}, 33(3), mar 2024.

\bibitem{harergan}
Jacob Harer, Onur Ozdemir, Tomo Lazovich, Christopher Reale, Rebecca Russell, Louis Kim, et~al.
\newblock Learning to repair software vulnerabilities with generative adversarial networks.
\newblock {\em Advances in neural information processing systems}, 31, 2018.

\bibitem{avrproblem1}
Kui Liu, Li~Li, Anil Koyuncu, Dongsun Kim, Zhe Liu, Jacques Klein, and Tegawendé~F. Bissyandé.
\newblock A critical review on the evaluation of automated program repair systems.
\newblock {\em Journal of Systems and Software}, 171:110817, 2021.

\bibitem{avrproblem2}
Manish Motwani, Mauricio Soto, Yuriy Brun, René Just, and Claire Le~Goues.
\newblock Quality of automated program repair on real-world defects.
\newblock {\em IEEE Transactions on Software Engineering}, 48(2):637--661, 2022.

\bibitem{avrproblem3}
Quanjun Zhang, Chunrong Fang, Yuxiang Ma, Weisong Sun, and Zhenyu Chen.
\newblock A survey of learning-based automated program repair.
\newblock {\em ACM Trans. Softw. Eng. Methodol.}, 33(2), dec 2023.

\bibitem{hoarelc}
C.A.R. Hoare.
\newblock Legacy code.
\newblock In {\em ICFEM 2000. Third IEEE International Conference on Formal Engineering Methods}, pages 75--75, 2000.

\bibitem{alexvulntime}
Nikolaos Alexopoulos, Manuel Brack, Jan~Philipp Wagner, Tim Grube, and Max M{\"u}hlh{\"a}user.
\newblock How long do vulnerabilities live in the code? a {Large-Scale} empirical measurement study on {FOSS} vulnerability lifetimes.
\newblock In {\em 31st USENIX Security Symposium (USENIX Security 22)}, pages 359--376, Boston, MA, August 2022. USENIX Association.

\bibitem{liucorrectness}
Jiawei Liu, Chunqiu~Steven Xia, Yuyao Wang, and LINGMING ZHANG.
\newblock Is your code generated by chatgpt really correct? rigorous evaluation of large language models for code generation.
\newblock In A.~Oh, T.~Neumann, A.~Globerson, K.~Saenko, M.~Hardt, and S.~Levine, editors, {\em Advances in Neural Information Processing Systems}, volume~36, pages 21558--21572. Curran Associates, Inc., 2023.

\bibitem{gpt4}
OpenAI.
\newblock Hello gpt-4o | openai, 2024.

\bibitem{llama3}
Meta Research.
\newblock Meta llama 3, 2024.

\bibitem{codellama}
Baptiste Rozière, Jonas Gehring, Fabian Gloeckle, Sten Sootla, Itai Gat, Xiaoqing~Ellen Tan, Yossi Adi, Jingyu Liu, Romain Sauvestre, Tal Remez, Jérémy Rapin, Artyom Kozhevnikov, Ivan Evtimov, Joanna Bitton, Manish Bhatt, Cristian~Canton Ferrer, Aaron Grattafiori, Wenhan Xiong, Alexandre Défossez, Jade Copet, Faisal Azhar, Hugo Touvron, Louis Martin, Nicolas Usunier, Thomas Scialom, and Gabriel Synnaeve.
\newblock Code llama: Open foundation models for code, 2023.

\bibitem{peft}
Haokun Liu, Derek Tam, Mohammed Muqeeth, Jay Mohta, Tenghao Huang, Mohit Bansal, and Colin~A Raffel.
\newblock Few-shot parameter-efficient fine-tuning is better and cheaper than in-context learning.
\newblock In S.~Koyejo, S.~Mohamed, A.~Agarwal, D.~Belgrave, K.~Cho, and A.~Oh, editors, {\em Advances in Neural Information Processing Systems}, volume~35, pages 1950--1965. Curran Associates, Inc., 2022.

\bibitem{lora}
Edward~J. Hu, Yelong Shen, Phillip Wallis, Zeyuan Allen-Zhu, Yuanzhi Li, Shean Wang, Lu~Wang, and Weizhu Chen.
\newblock Lora: Low-rank adaptation of large language models, 2021.

\bibitem{liu2022peft}
Haokun Liu, Derek Tam, Mohammed Muqeeth, Jay Mohta, Tenghao Huang, Mohit Bansal, and Colin Raffel.
\newblock Few-shot parameter-efficient fine-tuning is better and cheaper than in-context learning, 2022.

\bibitem{xiongfinetune}
Honglin Xiong, Sheng Wang, Yitao Zhu, Zihao Zhao, Yuxiao Liu, Linlin Huang, Qian Wang, and Dinggang Shen.
\newblock Doctorglm: Fine-tuning your chinese doctor is not a herculean task, 2023.

\bibitem{mistralmoe}
Albert~Q. Jiang, Alexandre Sablayrolles, Antoine Roux, Arthur Mensch, Blanche Savary, Chris Bamford, Devendra~Singh Chaplot, Diego de~las Casas, Emma~Bou Hanna, Florian Bressand, Gianna Lengyel, Guillaume Bour, Guillaume Lample, Lélio~Renard Lavaud, Lucile Saulnier, Marie-Anne Lachaux, Pierre Stock, Sandeep Subramanian, Sophia Yang, Szymon Antoniak, Teven~Le Scao, Théophile Gervet, Thibaut Lavril, Thomas Wang, Timothée Lacroix, and William~El Sayed.
\newblock Mixtral of experts, 2024.

\bibitem{prompteng1}
Yongchao Zhou, Andrei~Ioan Muresanu, Ziwen Han, Keiran Paster, Silviu Pitis, Harris Chan, and Jimmy Ba.
\newblock Large language models are human-level prompt engineers.
\newblock In {\em Advances in Neural Information Processing Systems}. Curran Associates, Inc., 2022.

\bibitem{prompteng2}
Hendrik Strobelt, Albert Webson, Victor Sanh, Benjamin Hoover, Johanna Beyer, Hanspeter Pfister, and Alexander~M Rush.
\newblock Interactive and visual prompt engineering for ad-hoc task adaptation with large language models.
\newblock {\em IEEE transactions on visualization and computer graphics}, 29(1):1146--1156, 2022.

\bibitem{prompteng3}
Paul Denny, Viraj Kumar, and Nasser Giacaman.
\newblock Conversing with copilot: Exploring prompt engineering for solving cs1 problems using natural language.
\newblock In {\em Proceedings of the 54th ACM Technical Symposium on Computer Science Education V. 1}, pages 1136--1142, 2023.

\bibitem{nvd}
National Vulnerability Database.
\newblock {\em NVD - Home}.

\bibitem{householdercwe}
Allen~D. Householder, Jeff Chrabaszcz, Trent Novelly, David Warren, and Jonathan~M. Spring.
\newblock Historical analysis of exploit availability timelines.
\newblock In {\em 13th USENIX Workshop on Cyber Security Experimentation and Test (CSET 20)}. USENIX Association, August 2020.

\bibitem{bugconcepts}
A.~Avizienis, J.-C. Laprie, B.~Randell, and C.~Landwehr.
\newblock Basic concepts and taxonomy of dependable and secure computing.
\newblock {\em IEEE Transactions on Dependable and Secure Computing}, 1(1):11--33, 2004.

\bibitem{monvulnrev}
Martin Monperrus.
\newblock Automatic software repair: A bibliography.
\newblock {\em ACM Comput. Surv.}, 51(1), jan 2018.

\bibitem{RAMPO}
Kohei Tsujio, Mohammad~Abdullah Al~Faruque, and Yasser Shoukry.
\newblock Rampo: A cegar-based integration of binary code analysis and system falsification for cyber-kinetic vulnerability detection.
\newblock In {\em 2024 ACM/IEEE 15th International Conference on Cyber-Physical Systems (ICCPS)}, pages 45--54, 2024.

\bibitem{zhengstaticdetect}
Yunhui Zheng and Xiangyu Zhang.
\newblock Path sensitive static analysis of web applications for remote code execution vulnerability detection.
\newblock In {\em 2013 35th International Conference on Software Engineering (ICSE)}, pages 652--661. IEEE, 2013.

\bibitem{limldetect}
Zhen Li, Deqing Zou, Shouhuai Xu, Xinyu Ou, Hai Jin, Sujuan Wang, Zhijun Deng, and Yuyi Zhong.
\newblock Vuldeepecker: {A} deep learning-based system for vulnerability detection.
\newblock In {\em 25th Annual Network and Distributed System Security Symposium, {NDSS} 2018, San Diego, California, USA, February 18-21, 2018}. The Internet Society, 2018.

\bibitem{chakramldetect}
Saikat Chakraborty, Rahul Krishna, Yangruibo Ding, and Baishakhi Ray.
\newblock Deep learning based vulnerability detection: Are we there yet?
\newblock {\em IEEE Transactions on Software Engineering}, 48(9):3280--3296, 2021.

\bibitem{sotirovdetect}
Alexander~Ivanov Sotirov.
\newblock {\em Automatic vulnerability detection using static source code analysis}.
\newblock PhD thesis, Citeseer, 2005.

\bibitem{xiallmrepair3}
Chunqiu~Steven Xia and Lingming Zhang.
\newblock Conversational automated program repair.
\newblock {\em arXiv preprint arXiv:2301.13246}, 2023.

\bibitem{fangsanitize}
Fang Yu, Ching-Yuan Shueh, Chun-Han Lin, Yu-Fang Chen, Bow-Yaw Wang, and Tevfik Bultan.
\newblock Optimal sanitization synthesis for web application vulnerability repair.
\newblock In {\em Proceedings of the 25th International Symposium on Software Testing and Analysis}, ISSTA 2016, page 189–200, New York, NY, USA, 2016. Association for Computing Machinery.

\bibitem{gaorepair}
Xiang Gao, Bo~Wang, Gregory~J Duck, Ruyi Ji, Yingfei Xiong, and Abhik Roychoudhury.
\newblock Beyond tests: Program vulnerability repair via crash constraint extraction.
\newblock {\em ACM Transactions on Software Engineering and Methodology (TOSEM)}, 30(2):1--27, 2021.

\bibitem{vcycle}
Natalie Paskoski.
\newblock Using the nist cybersecurity framework in your vulnerability management process.
\newblock Technical report, Retail and Hospitality Information Security and Analysis Center.

\bibitem{attention}
Ashish Vaswani, Noam Shazeer, Niki Parmar, Jakob Uszkoreit, Llion Jones, Aidan~N. Gomez, Lukasz Kaiser, and Illia Polosukhin.
\newblock Attention is all you need.
\newblock {\em CoRR}, abs/1706.03762, 2017.

\bibitem{shillmcontext}
Freda Shi, Xinyun Chen, Kanishka Misra, Nathan Scales, David Dohan, Ed~H Chi, Nathanael Sch{\"a}rli, and Denny Zhou.
\newblock Large language models can be easily distracted by irrelevant context.
\newblock In {\em International Conference on Machine Learning}, pages 31210--31227. PMLR, 2023.

\bibitem{zaheercontext}
Manzil Zaheer, Guru Guruganesh, Kumar~Avinava Dubey, Joshua Ainslie, Chris Alberti, Santiago Ontanon, Philip Pham, Anirudh Ravula, Qifan Wang, Li~Yang, et~al.
\newblock Big bird: Transformers for longer sequences.
\newblock {\em Advances in neural information processing systems}, 33:17283--17297, 2020.

\bibitem{lewisrag}
Patrick Lewis, Ethan Perez, Aleksandra Piktus, Fabio Petroni, Vladimir Karpukhin, Naman Goyal, Heinrich K{\"u}ttler, Mike Lewis, Wen-tau Yih, Tim Rockt{\"a}schel, et~al.
\newblock Retrieval-augmented generation for knowledge-intensive nlp tasks.
\newblock {\em Advances in Neural Information Processing Systems}, 33:9459--9474, 2020.

\bibitem{gpt3}
Tom Brown, Benjamin Mann, Nick Ryder, Melanie Subbiah, Jared~D Kaplan, Prafulla Dhariwal, Arvind Neelakantan, Pranav Shyam, Girish Sastry, Amanda Askell, et~al.
\newblock Language models are few-shot learners.
\newblock {\em Advances in neural information processing systems}, 33:1877--1901, 2020.

\bibitem{gpt4param}
{Maximilian Schreiner}.
\newblock Gpt-4 architecture, datasets, costs and more leaked.

\bibitem{llmemerge}
Jason Wei, Yi~Tay, Rishi Bommasani, Colin Raffel, Barret Zoph, Sebastian Borgeaud, Dani Yogatama, Maarten Bosma, Denny Zhou, Donald Metzler, et~al.
\newblock Emergent abilities of large language models.
\newblock {\em arXiv preprint arXiv:2206.07682}, 2022.

\bibitem{llmnoemerge}
Rylan Schaeffer, Brando Miranda, and Sanmi Koyejo.
\newblock Are emergent abilities of large language models a mirage?
\newblock {\em Advances in Neural Information Processing Systems}, 36, 2024.

\bibitem{gptpricing}
OpenAI.
\newblock Openai | pricing.

\bibitem{zakenpeft}
Elad~Ben Zaken, Shauli Ravfogel, and Yoav Goldberg.
\newblock Bitfit: Simple parameter-efficient fine-tuning for transformer-based masked language-models.
\newblock {\em arXiv preprint arXiv:2106.10199}, 2021.

\bibitem{liupeft}
Haokun Liu, Derek Tam, Mohammed Muqeeth, Jay Mohta, Tenghao Huang, Mohit Bansal, and Colin~A Raffel.
\newblock Few-shot parameter-efficient fine-tuning is better and cheaper than in-context learning.
\newblock {\em Advances in Neural Information Processing Systems}, 35:1950--1965, 2022.

\bibitem{fupeft}
Zihao Fu, Haoran Yang, Anthony Man-Cho So, Wai Lam, Lidong Bing, and Nigel Collier.
\newblock On the effectiveness of parameter-efficient fine-tuning.
\newblock In {\em Proceedings of the AAAI Conference on Artificial Intelligence}, volume~37, pages 12799--12807, 2023.

\bibitem{qlora}
Tim Dettmers, Artidoro Pagnoni, Ari Holtzman, and Luke Zettlemoyer.
\newblock Qlora: Efficient finetuning of quantized llms.
\newblock {\em Advances in Neural Information Processing Systems}, 36, 2024.

\bibitem{loramoe}
Wenfeng Feng, Chuzhan Hao, Yuewei Zhang, Yu~Han, and Hao Wang.
\newblock Mixture-of-loras: An efficient multitask tuning for large language models.
\newblock {\em arXiv preprint arXiv:2403.03432}, 2024.

\bibitem{zhoumoe}
Yanqi Zhou, Tao Lei, Hanxiao Liu, Nan Du, Yanping Huang, Vincent Zhao, Andrew~M Dai, Quoc~V Le, James Laudon, et~al.
\newblock Mixture-of-experts with expert choice routing.
\newblock {\em Advances in Neural Information Processing Systems}, 35:7103--7114, 2022.

\bibitem{dumoescale}
Nan Du, Yanping Huang, Andrew~M Dai, Simon Tong, Dmitry Lepikhin, Yuanzhong Xu, Maxim Krikun, Yanqi Zhou, Adams~Wei Yu, Orhan Firat, et~al.
\newblock Glam: Efficient scaling of language models with mixture-of-experts.
\newblock In {\em International Conference on Machine Learning}, pages 5547--5569. PMLR, 2022.

\bibitem{artetxemoeperf}
Mikel Artetxe, Shruti Bhosale, Naman Goyal, Todor Mihaylov, Myle Ott, Sam Shleifer, Xi~Victoria Lin, Jingfei Du, Srinivasan Iyer, Ramakanth Pasunuru, et~al.
\newblock Efficient large scale language modeling with mixtures of experts.
\newblock {\em arXiv preprint arXiv:2112.10684}, 2021.

\bibitem{shtedpe}
Aleksandar Shtedritski, Christian Rupprecht, and Andrea Vedaldi.
\newblock What does clip know about a red circle? visual prompt engineering for vlms.
\newblock In {\em Proceedings of the IEEE/CVF International Conference on Computer Vision}, pages 11987--11997, 2023.

\bibitem{whitepe}
Jules White, Quchen Fu, Sam Hays, Michael Sandborn, Carlos Olea, Henry Gilbert, Ashraf Elnashar, Jesse Spencer-Smith, and Douglas~C Schmidt.
\newblock A prompt pattern catalog to enhance prompt engineering with chatgpt.
\newblock {\em arXiv preprint arXiv:2302.11382}, 2023.

\bibitem{chainofthought}
Jason Wei, Xuezhi Wang, Dale Schuurmans, Maarten Bosma, Fei Xia, Ed~Chi, Quoc~V Le, Denny Zhou, et~al.
\newblock Chain-of-thought prompting elicits reasoning in large language models.
\newblock {\em Advances in neural information processing systems}, 35:24824--24837, 2022.

\bibitem{kojimazeroshot}
Takeshi Kojima, Shixiang~Shane Gu, Machel Reid, Yutaka Matsuo, and Yusuke Iwasawa.
\newblock Large language models are zero-shot reasoners.
\newblock {\em Advances in neural information processing systems}, 35:22199--22213, 2022.

\bibitem{kliebercompiler}
William Klieber, Ruben Martins, Ryan Steele, Matt Churilla, Mike McCall, and David Svoboda.
\newblock Automated code repair to ensure spatial memory safety.
\newblock In {\em 2021 IEEE/ACM International Workshop on Automated Program Repair (APR)}, pages 23--30. IEEE, 2021.

\bibitem{landmanstatic}
Davy Landman, Alexander Serebrenik, and Jurgen~J Vinju.
\newblock Challenges for static analysis of java reflection-literature review and empirical study.
\newblock In {\em 2017 IEEE/ACM 39th International Conference on Software Engineering (ICSE)}, pages 507--518. IEEE, 2017.

\bibitem{chentransfer}
Zimin Chen, Steve Kommrusch, and Martin Monperrus.
\newblock Neural transfer learning for repairing security vulnerabilities in c code.
\newblock {\em IEEE Transactions on Software Engineering}, 49(1):147--165, 2022.

\bibitem{codet5}
Yue Wang, Weishi Wang, Shafiq Joty, and Steven~CH Hoi.
\newblock Codet5: Identifier-aware unified pre-trained encoder-decoder models for code understanding and generation.
\newblock {\em arXiv preprint arXiv:2109.00859}, 2021.

\bibitem{CREAM}
Quanjun Zhang, Chunrong Fang, Ye~Shang, Tongke Zhang, Shengcheng Yu, and Zhenyu Chen.
\newblock No man is an island: Towards fully automatic programming by code search, code generation and program repair, 2024.

\bibitem{llmfix}
Hao Wen, Yueheng Zhu, Chao Liu, Xiaoxue Ren, Weiwei Du, and Meng Yan.
\newblock Fixing code generation errors for large language models, 2024.

\bibitem{peillmreason}
Kexin Pei, David Bieber, Kensen Shi, Charles Sutton, and Pengcheng Yin.
\newblock Can large language models reason about program invariants?
\newblock In {\em International Conference on Machine Learning}, pages 27496--27520. PMLR, 2023.

\bibitem{codebert}
Zhangyin Feng, Daya Guo, Duyu Tang, Nan Duan, Xiaocheng Feng, Ming Gong, Linjun Shou, Bing Qin, Ting Liu, Daxin Jiang, et~al.
\newblock Codebert: A pre-trained model for programming and natural languages.
\newblock {\em arXiv preprint arXiv:2002.08155}, 2020.

\bibitem{codex}
Mark Chen, Jerry Tworek, Heewoo Jun, Qiming Yuan, Henrique Ponde de~Oliveira Pinto, Jared Kaplan, Harri Edwards, Yuri Burda, Nicholas Joseph, Greg Brockman, et~al.
\newblock Evaluating large language models trained on code.
\newblock {\em arXiv preprint arXiv:2107.03374}, 2021.

\bibitem{wizardcoder}
Ziyang Luo, Can Xu, Pu~Zhao, Qingfeng Sun, Xiubo Geng, Wenxiang Hu, Chongyang Tao, Jing Ma, Qingwei Lin, and Daxin Jiang.
\newblock Wizardcoder: Empowering code large language models with evol-instruct.
\newblock {\em arXiv preprint arXiv:2306.08568}, 2023.

\bibitem{codegen}
Erik Nijkamp, Bo~Pang, Hiroaki Hayashi, Lifu Tu, Huan Wang, Yingbo Zhou, Silvio Savarese, and Caiming Xiong.
\newblock Codegen: An open large language model for code with multi-turn program synthesis.
\newblock {\em arXiv preprint arXiv:2203.13474}, 2022.

\bibitem{safe}
Van Nguyen, Surya Nepal, Tingmin Wu, Xingliang Yuan, and Carsten Rudolph.
\newblock Safe: Advancing large language models in leveraging semantic and syntactic relationships for software vulnerability detection, 2024.

\bibitem{autosafecoder}
Ana Nunez, Nafis~Tanveer Islam, Sumit~Kumar Jha, and Peyman Najafirad.
\newblock Autosafecoder: A multi-agent framework for securing llm code generation through static analysis and fuzz testing, 2024.

\bibitem{xullmeval}
Frank~F Xu, Uri Alon, Graham Neubig, and Vincent~Josua Hellendoorn.
\newblock A systematic evaluation of large language models of code.
\newblock In {\em Proceedings of the 6th ACM SIGPLAN International Symposium on Machine Programming}, pages 1--10, 2022.

\bibitem{llm4plc}
Mohamad Fakih, Rahul Dharmaji, Yasamin Moghaddas, Gustavo Quiros, Oluwatosin Ogundare, and Mohammad~Abdullah Al~Faruque.
\newblock Llm4plc: Harnessing large language models for verifiable programming of plcs in industrial control systems.
\newblock In {\em Proceedings of the 46th International Conference on Software Engineering: Software Engineering in Practice}, ICSE-SEIP '24, page 192–203, New York, NY, USA, 2024. Association for Computing Machinery.

\bibitem{yangllmrepair}
Boyang Yang, Haoye Tian, Jiadong Ren, Hongyu Zhang, Jacques Klein, Tegawend{\'e}~F Bissyand{\'e}, Claire~Le Goues, and Shunfu Jin.
\newblock Multi-objective fine-tuning for enhanced program repair with llms.
\newblock {\em arXiv preprint arXiv:2404.12636}, 2024.

\bibitem{silvallmrepair}
Andr{\'e} Silva, Sen Fang, and Martin Monperrus.
\newblock Repairllama: Efficient representations and fine-tuned adapters for program repair.
\newblock {\em arXiv preprint arXiv:2312.15698}, 2023.

\bibitem{zhoullmrepairreview}
Xin Zhou, Sicong Cao, Xiaobing Sun, and David Lo.
\newblock Large language model for vulnerability detection and repair: Literature review and roadmap.
\newblock {\em arXiv preprint arXiv:2404.02525}, 2024.

\bibitem{jinllmrepair}
Matthew Jin, Syed Shahriar, Michele Tufano, Xin Shi, Shuai Lu, Neel Sundaresan, and Alexey Svyatkovskiy.
\newblock Inferfix: End-to-end program repair with llms.
\newblock In {\em Proceedings of the 31st ACM Joint European Software Engineering Conference and Symposium on the Foundations of Software Engineering}, ESEC/FSE 2023, page 1646–1656, New York, NY, USA, 2023. Association for Computing Machinery.

\bibitem{xiallmrepair}
Chunqiu~Steven Xia, Yuxiang Wei, and Lingming Zhang.
\newblock Automated program repair in the era of large pre-trained language models.
\newblock In {\em 2023 IEEE/ACM 45th International Conference on Software Engineering (ICSE)}, pages 1482--1494, 2023.

\bibitem{joshillmrepair}
Harshit Joshi, Jos{\'e}~Cambronero Sanchez, Sumit Gulwani, Vu~Le, Gust Verbruggen, and Ivan Radi{\v{c}}ek.
\newblock Repair is nearly generation: Multilingual program repair with llms.
\newblock In {\em Proceedings of the AAAI Conference on Artificial Intelligence}, volume~37, pages 5131--5140, 2023.

\bibitem{ahmedllmrepair}
Toufique Ahmed and Premkumar Devanbu.
\newblock Better patching using llm prompting, via self-consistency.
\newblock In {\em 2023 38th IEEE/ACM International Conference on Automated Software Engineering (ASE)}, pages 1742--1746, 2023.

\bibitem{xiallmrepair2}
Chunqiu~Steven Xia and Lingming Zhang.
\newblock Keep the conversation going: Fixing 162 out of 337 bugs for \$0.42 each using chatgpt.
\newblock {\em arXiv preprint arXiv:2304.00385}, 2023.

\bibitem{pearcezeroshot}
Hammond Pearce, Benjamin Tan, Baleegh Ahmad, Ramesh Karri, and Brendan Dolan-Gavitt.
\newblock Examining zero-shot vulnerability repair with large language models.
\newblock In {\em 2023 IEEE Symposium on Security and Privacy (SP)}, pages 2339--2356, 2023.

\bibitem{wengllmrepair}
Guoyang Weng and Artur Andrzejak.
\newblock Automatic bug fixing via deliberate problem solving with large language models.
\newblock In {\em 2023 IEEE 34th International Symposium on Software Reliability Engineering Workshops (ISSREW)}, pages 34--36, 2023.

\bibitem{jainllmrepair}
Naman Jain, Skanda Vaidyanath, Arun Iyer, Nagarajan Natarajan, Suresh Parthasarathy, Sriram Rajamani, and Rahul Sharma.
\newblock Jigsaw: large language models meet program synthesis.
\newblock In {\em Proceedings of the 44th International Conference on Software Engineering}, ICSE '22, page 1219–1231, New York, NY, USA, 2022. Association for Computing Machinery.

\bibitem{syslang}
James~J. Horning.
\newblock Yes! high level languages should be used to write systems software.
\newblock In {\em Proceedings of the 1975 Annual Conference}, ACM '75, page 206–208, New York, NY, USA, 1975. Association for Computing Machinery.

\bibitem{syslang2}
James~S. Rogers.
\newblock Language choice for safety critical applications.
\newblock {\em Ada Lett.}, 31(3):81–90, nov 2011.

\bibitem{codebleu}
Shuo Ren, Daya Guo, Shuai Lu, Long Zhou, Shujie Liu, Duyu Tang, Neel Sundaresan, Ming Zhou, Ambrosio Blanco, and Shuai Ma.
\newblock Codebleu: a method for automatic evaluation of code synthesis.
\newblock {\em arXiv preprint arXiv:2009.10297}, 2020.

\bibitem{bleu}
Kishore Papineni, Salim Roukos, Todd Ward, and Wei-Jing Zhu.
\newblock Bleu: a method for automatic evaluation of machine translation.
\newblock In {\em Proceedings of the 40th annual meeting of the Association for Computational Linguistics}, pages 311--318, 2002.

\bibitem{cvefixes}
Guru Bhandari, Amara Naseer, and Leon Moonen.
\newblock Cvefixes: automated collection of vulnerabilities and their fixes from open-source software.
\newblock In {\em Proceedings of the 17th International Conference on Predictive Models and Data Analytics in Software Engineering}, PROMISE 2021, page 30–39, New York, NY, USA, 2021. Association for Computing Machinery.

\bibitem{gptdocs}
OpenAI.
\newblock Models - openai api.

\bibitem{mobugtime}
Seyed Mohammadjavad~Seyed Talebi, Zhihao Yao, Ardalan~Amiri Sani, Zhiyun Qian, and Daniel Austin.
\newblock Undo workarounds for kernel bugs.
\newblock In {\em 30th USENIX Security Symposium (USENIX Security 21)}, pages 2381--2398. USENIX Association, August 2021.

\end{thebibliography}

\end{document}